\documentclass[epj]{svjour}

\usepackage{graphics}

\begin{document}

\title{A model of coupled maps with Pareto behavior}
\author{J. R. S\'{a}nchez\inst{1}, J. Gonz\'{a}lez-Est\'{e}vez\inst{2,}\inst{4} 
\thanks{\emph{E-mail address:} jgonzale@unet.edu.ve},
R. L\'{o}pez-Ruiz\inst{3},  and M. G. Cosenza\inst{4}
}                     

\institute{Facultad de Ingenier\'{\i}a, Universidad Nacional de Mar del Plata, Av. J. B. Justo 4302, 7600-Mar del Plata, Argentina. \and
Laboratorio de F\'{\i}sica Aplicada y Computacional, Universidad Nacional Experimental del T\'{a}chira, San Crist\'{o}bal, Venezuela. \and 
Facultad de Ciencias, DIIS and BIFI, Universidad de Zaragoza, 50009-Zaragoza, Spain. \and 
Centro de F\'{\i}sica Fundamental, Universidad de Los Andes, M\'{e}rida, Apartado Postal 26, M\'{e}rida 5251, Venezuela.}

\date{Received: date / Revised version: date}

\abstract{
A deterministic system of coupled maps is proposed as a model for economic activity among interacting agents. The values of the maps represent the wealth of the agents. The dynamics of the system is controlled by two parameters. One parameter expresses the growth capacity of the agents and the other describes the local environmental pressure. For some values of the parameters, the system exhibits nontrivial collective behavior, characterized by macroscopic periodic oscillations of the average wealth of the system, emerging out of local chaos. The probability distribution of wealth in the asymptotic regime shows a power law behavior for some ranges of parameters.
\PACS{{05.45.-a}{} \and {05.45.Ra}{} \and {89.65.-s}{} \and {89.65.Gh}{}
     } 
} 

\titlerunning{A Model of Coupled-Maps for Economic Dynamics}
\authorrunning{S\'{a}nchez, Gonz\'{a}lez-Est\'{e}vez, L\'{o}pez-Ruiz, Cosenza}

\maketitle

The study of wealth distribution in western societies has been a focus of much attention in the emerging area of research of Econophysics. A power law behavior is found in many economic activities,
for instance, in income distributions. Typically, high income earners, amounting to a few percent of the population, are distributed following a Pareto-like distribution. \cite{pareto,lorenz,raw,klass,reed}. Hence, inequality in the wealth distribution is a fact in most economic activities. The origin of such behavior seems to be caused by the interaction of the macro with the microeconomy. Here we propose a simple deterministic spatiotemporal model for economic dynamics where inequality emerges as a result of the dynamical processes taking place only at the microscopic scale.
That is, the microeconomy fully determines the macroeconomic characteristics of the system.

The model \cite{sanchez} consists of $N$ interacting agents representing companies, countries or other economic entities, placed as nodes on a network. For simplicity, we shall assume that the agents are distributed on a one-dimensional lattice with periodic boundary conditions. The state of an agent $i$, $i=1 \cdots N$, is characterized by a real variable $x_t(i) \in [0,\infty]$ denoting its wealth or richness at the discrete time $t$. 
The system evolves in time synchronously. Each agent updates its state $x_t(i)$ according to its present state and the states of its nearest neighbors. Thus, in a first approach,
we propose that the value of $x_{t+1}(i)$ is given by the product of two terms; the {\it natural growth} of agent $i$ given by $r(i)x_{t}(i)$ with positive local ratio $r(i)$, and a {\it control term} that limits this growth with respect to the local field
\begin{equation}
\Psi_t(i)=\frac{1}{2}\left[ x_{t}(i-1)+x_{t}(i+1)\right] ,
\label{field}
\end{equation}
through a negative exponential with parameter $a(i)$,
\begin{equation}
x_{t+1}(i) = r(i)\:x_{t}(i)\: \exp(-\mid x_{t}(i)-a(i)\Psi_t(i)\mid).
\label{1dcml}
\end{equation}
The parameter $r(i)$ represents the {\it capacity} of agent $i$ to get richer
and the parameter $a(i)$ describes the local {\it selection pressure} \cite{ausloos}.
This means that the largest rate of growth for agent $i$ is obtained
when $x(i)\simeq a(i)\Psi_t(i)$, i.e., when the agent has reached some kind of adaptation 
to the local environment. In this paper we consider a homogeneous system with a uniform capacity $r$ and a fixed selection pressure $a$ for all the agents.

The simplest collective behavior of the one-dimensional coupled map system described by Eqs. (\ref{field}) and (\ref{1dcml}) corresponds to the synchronized or spatially uniform state. This state satisfies $x_t(i)=x_t$ and $\Psi_t(i)=x_t$,
$\forall i$, and its evolution is determined by the single map
\begin{equation}
x_{t+1}=r x_{t} \exp(-\mid (1-a)x_{t}\mid).
\label{map}
\end{equation}
For $r<1$ the synchronized dynamics Eq.~(\ref{map}) relaxes to the stable fixed point $x=0$, while for $r>1$ the synchronized system displays a sequence of bifurcations through different periodic and chaotic attractors, except for the singular case $a=1$. For $r>1$ the stable fixed point is $x=\log r / \left| 1-a \right|$. This point becomes unstable by a flip bifurcation at $r=e^2$. For increasing $r$, the entire period-doubling cascade characteristic of unimodal maps and other complex dynamical behaviors are generated by 
Eq.~(\ref{map}). However, it can be shown that such synchronized states are unstable. When a perturbation is introduced in the initial uniform state, the asymptotic dynamical state of the system is found to be more complex.

\begin{figure}[htb] 
\centerline{\rotatebox{0}{\resizebox{0.4\textwidth}{!}{\includegraphics{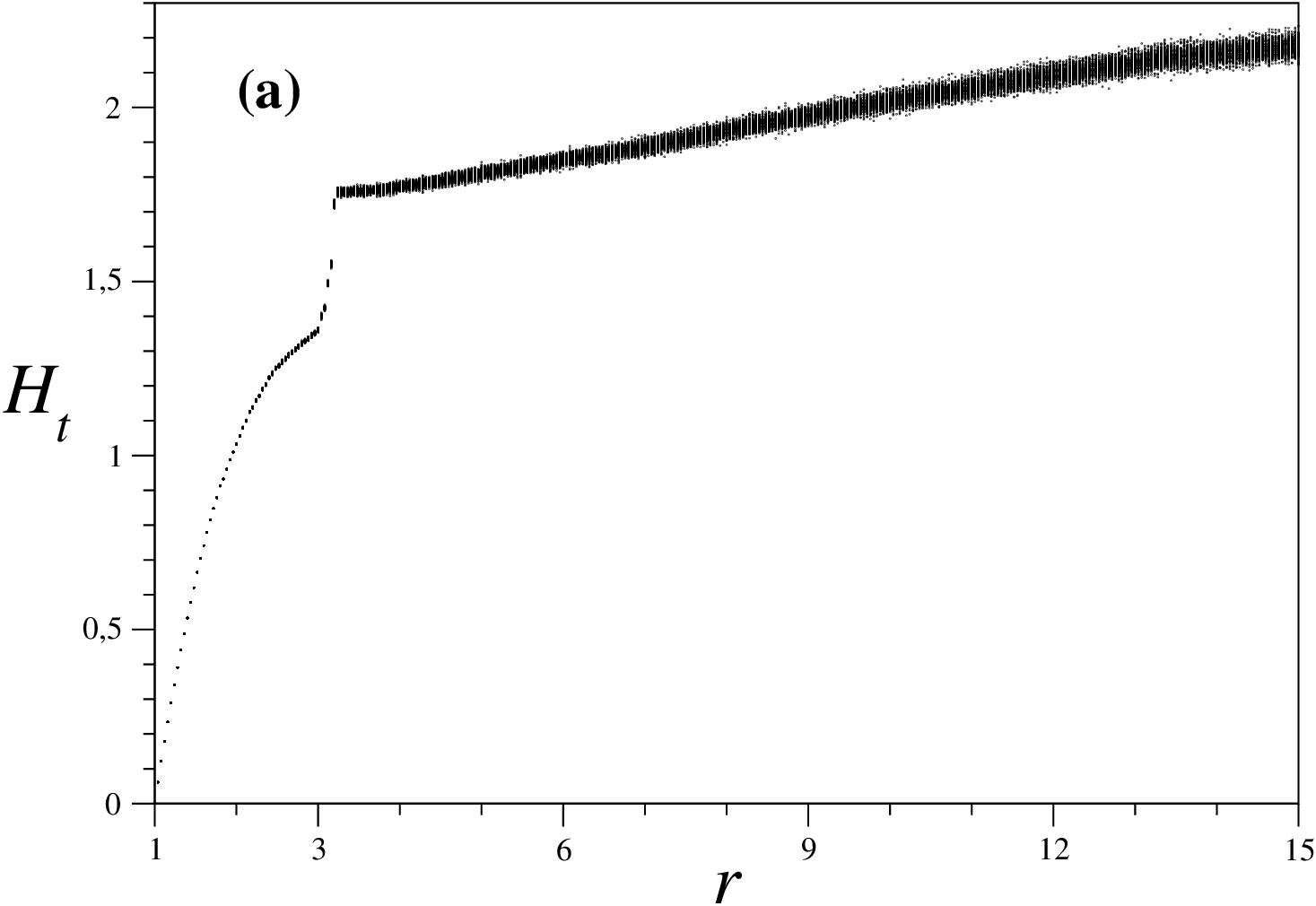}}}}
\centerline{\rotatebox{0}{\resizebox{0.4\textwidth}{!}{\includegraphics{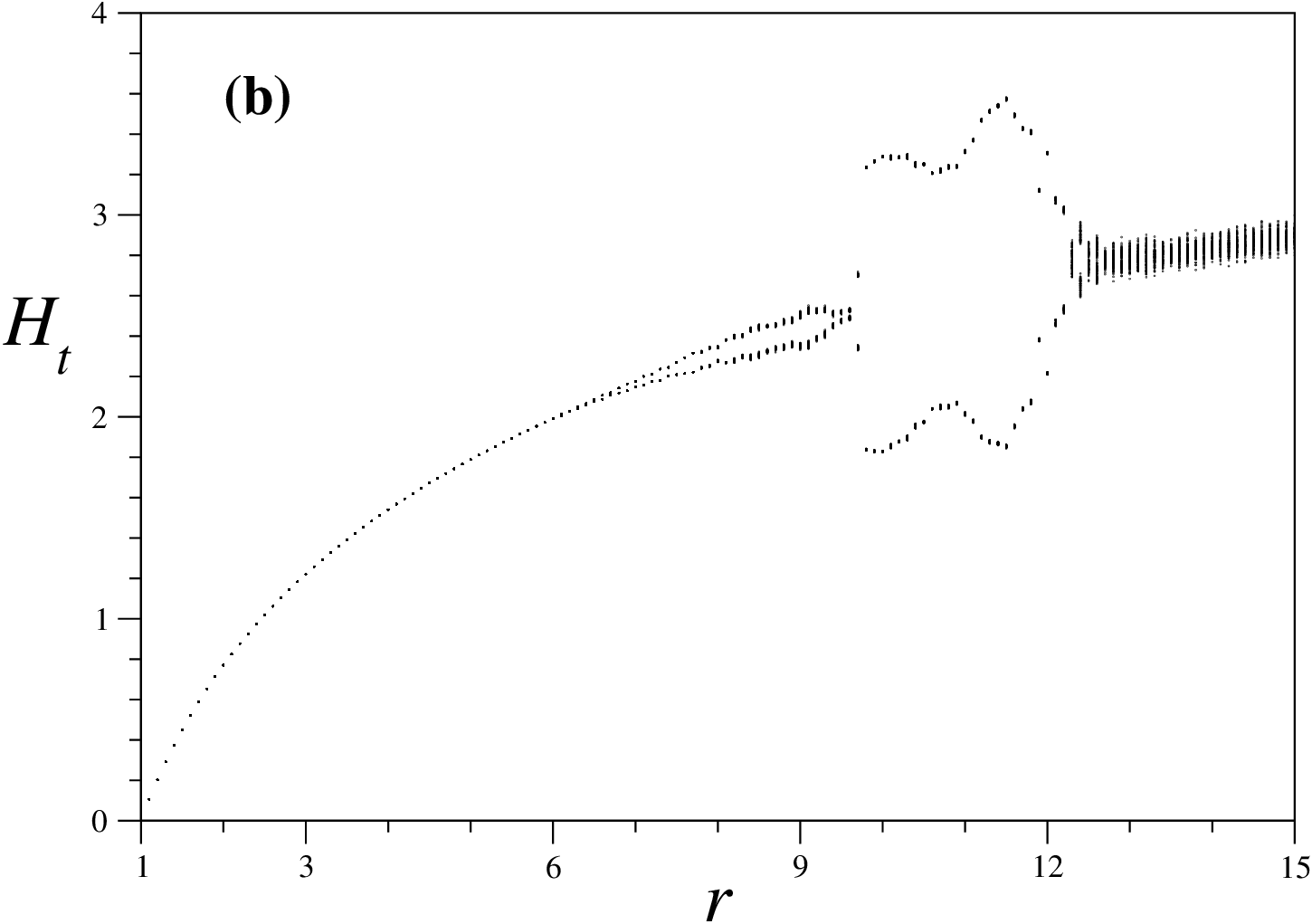}}}}
\caption{\em $H_{t}$ as a function of $r$, for two values of $a$. For each value of $r$, $H_{t}$ is plotted for $100$ iterations, after discarding $9900$ transients. System size $N = 10^{4}$. (a) $a=0.67$. (b) $a=0.1$.
\label{fig1a}}
\end{figure}

In general, the collective behavior of the system can be characterized through the instantaneous mean field of the network, defined as
\begin{equation}
H_{t}=\frac{1}{N}\sum_{j=1}^{N}x_{t}(j).
\label{global}
\end{equation}
Figures \ref{fig1a}(a) and \ref{fig1a}(b) show a bifurcation diagram of $H_{t}$ as a function of the parameter $r$, for two different values of the selection pressure $a$. In Fig.~\ref{fig1a}(a) it can be seen that $H_{t}$ reaches stationary values with some intrinsic fluctuations due to the local chaotic dynamics. However, for some parameter values, nontrivial collective behavior \cite{chate} can arise in this system as shown in Fig.~\ref{fig1a}(b). In this case, a macroscopic variable such as the mean field or average wealth in the system follows a periodic behavior coexisting with chaos at the microscopic level.

The statistical properties of the system can be expressed in terms of the probability distribution of wealth among the agents. For some values of the parameters $r$ and $a$, the distribution of wealth displays a power law behavior, as shown in Fig.~2. For the values of parameters chosen, the probability distribution scales as $P(x) \sim s^{\:\alpha}$, with $\alpha = -2.86$, similarly to scaling behaviors of Pareto type directly obtained from actual economy data \cite{nunes}.

\begin{figure}[htb] 
\centerline{\rotatebox{0}{\resizebox{0.4\textwidth}{!}{\includegraphics{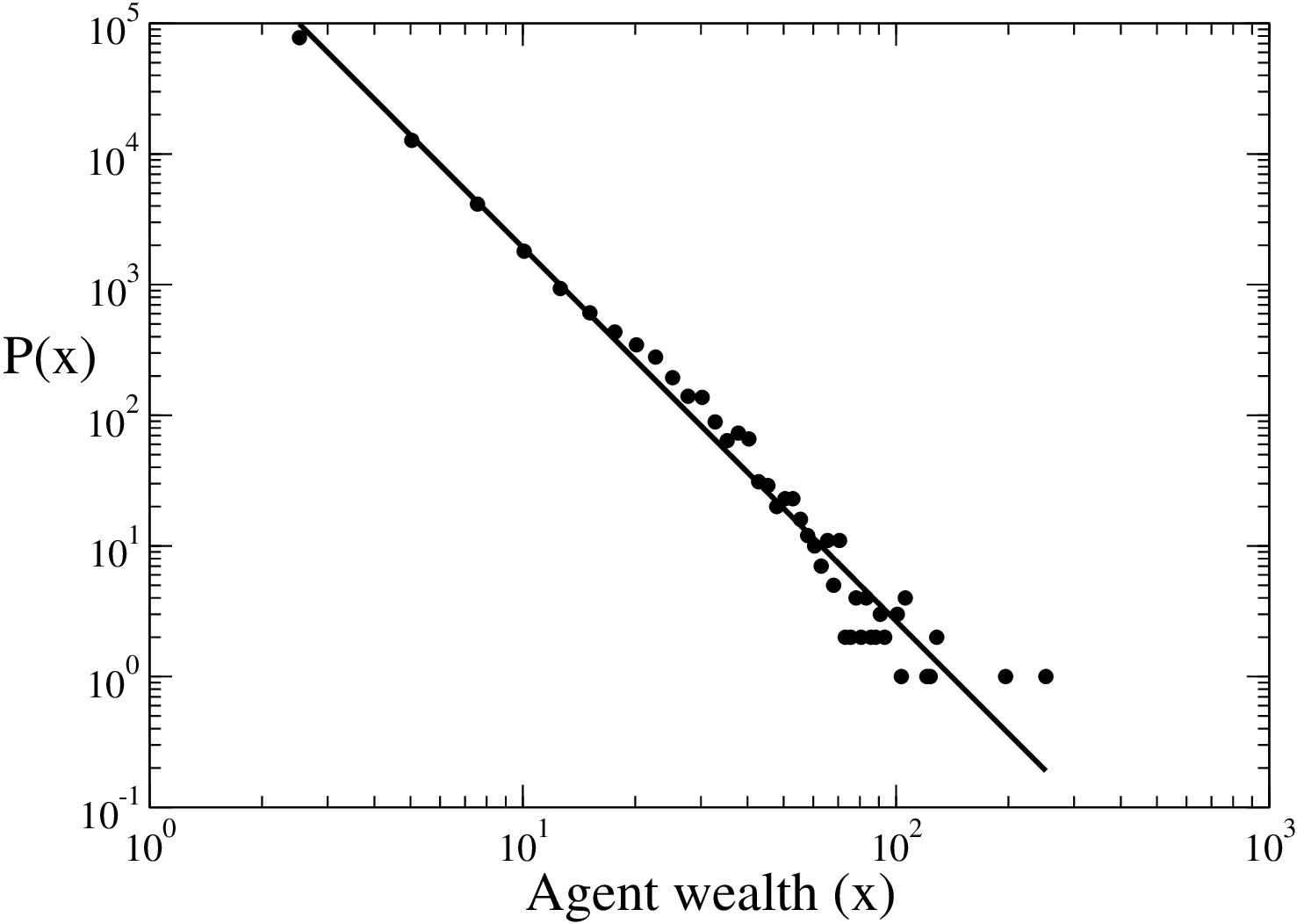}}}}
\caption{\em Log-log plot of the asymptotic probability distribution of wealth $P(x)$ vs. agent wealth $x$. The distribution $P(x)$ is calculated averaging the outcomes at $t=10^{4}$ of $100$ realizations of random initial conditions. Parameter values are $r=12$, $a=0.67$, $N=10^{4}$. The slope obtained is $\alpha = -2.86$ with a correlation coefficient $\beta = 0.98$.
\label{fig2}}
\end{figure}

In summary, we have shown that a power law behavior in the probability distribution of wealth can arise for some values of parameters in a deterministic system of interacting economic agents, such as in the coupled map model considered here. In addition, nontrivial collective behavior, where macroscopic order coexists with local disorder, can emerge in this system.

This work was supported in part by Decanato de Investigaci\'on of Universidad Nacional Experimental del T\'achira (UNET) and by FONACIT, Venezuela, under grants 04-001-2006 and F-2002000426, respectively. J.G.E. thanks Decanato de Investigaci\'on and Vicerrectorado Acad\'emico of UNET for travel support to the Universidad de Zaragoza, Spain. R. L.-R. acknowledges some financial support from the spanish research project FIS2004-05073-C04-01.

\end{document}